\documentclass[fleqn,10pt]{wlscirep}
\usepackage[utf8]{inputenc}
\usepackage[T1]{fontenc}

\def\be{\begin{equation}}
\def\ee{\end{equation}}
\def\beq{\begin{eqnarray}}
\def\eeq{\end{eqnarray}}
\def\n{\nonumber}

\usepackage{graphicx}
\usepackage{dcolumn}
\usepackage{amsmath}

\title{Strange stars in the framework of higher curvature
gravity}

\author[1,+]{Sudan Hansraj}
\author[2,*]{Megandhren Govender}
\author[1,**]{Lushen Moodly}
\author[3,4,$\dagger$]{Ksh. Newton Singh}
\affil[1]{Astrophysics Research Centre, School of Mathematics, Statistics and Computer Science, University of KwaZulu--Natal, Private Bag X54001, Durban 4000, South Africa }
\affil[2]{Department of Mathematics, Steve Biko Campus, Durban University of Technology, Durban 4000, South Africa}
\affil[3]{Department of Physics, National Defence Academy Khadakwasla,Pune-411023, India.}
\affil[4]{Department of Mathematics, Jadavpur University, Kolkata-700032, India.}

\affil[+]{hansrajs@ukzn.ac.za}
\affil[*]{megandhreng@dut.ac.za}
\affil[**]{lmoodly@gmail.com}
\affil[$\dagger$]{ntnphy@gmail.com}

\begin{abstract}
We study the influence of higher curvature effects on stellar structure and conclude that the properties of stars are greatly impacted when such terms are dynamic. In particular the surface gravitational redshift which is connected to the equation of state and also the mass-radius ratio differs greatly from the corresponding values in general relativity as evidenced through our empirical comparisons.  A model of a superdense star with strange star equation of state is constructed within the framework of the  Einstein--Gauss--Bonnet theory. Under these assumptions large classes of solutions are admitted by the field equations. We isolate a particular class with the ansatz of the Vaidya--Tikekar superdense star spatial gravitational potential. The model is found to satisfy elementary requirements for physical applicability and stability.   The parameter values chosen are consistent with observed star models. A significant effect of the higher curvature terms is to reduce the speed of sound and to drastically reduce the values of the surface gravitational redshift compared to the Einstein counterpart. These latter results have implications for interpretations of observations in relativistic astrophysics which are often made against the background of the standard general theory of relativity.
\end{abstract}
\begin{document}

\flushbottom
\maketitle
%
%
\thispagestyle{empty}


\section*{Introduction}

Alternate or extended theories of gravity have aroused considerable interest recently in view of difficulties with the general theory of relativity to explain anomalous behavior of gravitational phenomena such as the late time accelerated expansion of the universe (Riess et al. 1998; Perlmutter et al. 1999). One attempt at resolving the problem involves conjecturing the existence of exotic matter fields such as dark matter, dark energy, phantom fields and quintessence fields to name a few. To date there exists no experimental support for these ideas however a number of experiments are ongoing. In order to explain dark energy and dark matter de Rham (2014) suggests that the graviton is not massless but actually carries a small mass. This itself has a number of ramifications for physics which has been dealt with elsewhere in the literature.   An alternative approach is to re-examine the geometrical side of the field equations. Higher curvature effects may have a role to play.  In particular, Einstein--Gauss--Bonnet (EGB) theory has proved promising in this regard, and therefore is extensively studied. Note that EGB belongs to a more general class of theories called the Lovelock polynomial Lagrangians which constitute the most general tensor theory generating at most second order equations of motion. If the Lagrangian is allowed to involve both tensor and scalar fields then the most general such theory is due to Horndeski (1974). A further strong motivation for EGB theory is that the Gauss--Bonnet Lagrangian appears in a natural way in the effective action of heterotic string theory in the low energy limit (Gross 1999).  The causal structure of the singularities is different from general relativity for inhomogeneous distributions of dust and null dust (Ghosh, Jhingan \& Deshkar 2014). The question we are probing is whether the addition of higher curvature gravitational effects play a significant role in the evolution of stars. Indeed most of our understanding of observations in relativistic astrophysics are made on the basis of Einstein's theory of general relativity (GR). However, if GR is to be superseded by a higher curvature theory which preserves second order equations of motion and which reduces to GR in the solar system scale limit, then it is natural to ask what effect the higher curvature contributions have on stellar structure.

Granted five dimensional stars are not physically accessible however their existence has not been ruled out. The earliest work on higher dimensional gravity originated with Kaluza (1921) and Klein (1926) who considered a 5 dimensional manifold and ascribed the behavior of the electrodynamical field to four components of the metric tensor, ten to the usual four dimensional spacetime manifold and an extra dimension to a scalar field. Subsequently modern works on brane world gravity necessitating higher dimensions also proceeded. A comprehensive review was compiled by Maartens and Koyama (2010). The customary explanation for extra dimensions is that these are topologically curled and of very small size. Note that the Large Hadron Collider experiment also searched for extra dimensions but of large scale but did not detect any. This however, does not eliminate the existence of extra dimensions at microscopic scale. Indeed, spacetime dimensions of the size of 10 and 11 are essential  in quantum field theory.  While their magnitude is small, their effect on aspects of the gravitational field may be of immense importance as we show in this article.

Black holes in the context of  EGB theory have been rigorously studied in the literature. The higher dimensional Einstein models  of Tangherlini (1963) and Myers \& Perry (1986) were generalized to the the EGB regime of higher curvature gravity in the classic paper of Boulware \& Deser (1985). Further treatments of black holes in EGB are attributed to the works of Wheeler (1986), Myers \& Simons (1988) and Torii \& Maeda (2005). Maeda (2006) investigated the inhomogeneous collapse of dust however the exact solutions for the five dimensional case by Jhingan and Ghosh (2010) revealed that the collapse led to the formation of a massive but weak timelike singularity in contrast with general relativity. The configuration of an incompressible (constant density) hypersphere was considered by Dadhich et al (2010) who showed that the usual Schwarzschild interior solution of four dimensional gravity still holds in the higher dimensional and higher curvature arena.

 The universality of the Schwarzschild solution by  Dadhich et al. (2010) constituted the first nontrivial perfect fluid stellar model in EGB gravity. However, like its four dimensional counterpart, the solution continues to inherit the pathology of an infinite speed of sound rendering the metric physically unreasonable.  Kang et al. (2012) devised a static model of a star however there were two problems with the construction. Firstly, the solution still required a further integration to be completed in order to unlock the full flavor of the metric. This arose primarily because the continuity equation was being used and in the standard theory such an excursion almost certainly ends up requiring numerical integration.  Secondly, in view of the incomplete solution, it was not possible to match the solution with the exterior Boulware-Deser metric. Notwithstanding these matters, it is indeed interesting that Kang et al (2012) were able to obtain the Boulware-Deser solution in the limit of vanishing pressure and energy density.  It should be observed that boundary conditions for EGB gravity were derived in general form by Davis (2003) and the consequences are expected to be different from general relativity. For example, it is still not known if the matching of the first and second fundamental forms is equivalent to the existence of a vanishing pressure-free hypersurface as is the case in general relativity.  Variable density spherically symmetric exact solutions to the EGB field equations were first obtained by Hansraj et al. (2015), Chilambwe et al. (2015) and Maharaj et al. (2015) and shown to be consistent with the usual elementary expectations of astrophysical models.

 Local anisotropy in self-gravitating systems has been extensively studied within the framework of classical general relativity (Govender \& Thirukkanesh 2015; Maurya, Ratanpal \& Govender 2017; Maurya \& Govender 2017; Maurya et al. 2019). The inclusion of pressure anisotropy in the study of compact objects such as pulsars, neutrons stars and quark stars in 4-D gravity has led to physically viable stellar models. Analyses of the physical attributes of these models such as density profiles, pressure profiles, compactness and surface redshift agree with observed data within experimental error. The anisotropy parameter $\Delta = p_T - p_R$, where $p_T$ refers to the tangential pressure and $p_R$ is the radial pressure,  can either be positive or negative at each interior point of the matter configuration. When $p_T > p_R$ the force due to local anisotropy is repulsive which may lead to more massive and stable configurations. A fruitful approach to generating physically realizable models of compact objects is to consider the idea of embedding a 4-dimensional spherically symmetric spacetime into a 5-dimensional Euclidean space. Karmarkar (1948) classified these spacetimes as embedding class-1. It is interesting to note that the Karmarkar condition in the presence of {\em pressure isotropy} leads to two exact models. The Kohler-Chao (1965) solution can be interpreted as of cosmological nature as there is no finite radius for which the radial pressure vanishes. The other exact model is the interior Schwarzschild solution which suffers from various pathologies such as superluminal sound speeds within the stellar core. Bowers \& Liang (1974) demonstrated that the surface redshift can be arbitrarily large in the presence of pressure anisotropy. They were able to show that if the fractional anisotropy, $\frac{p_T - p_R}{p} > 0$, then the associated surface redshift is greater than its isotropic counterpart. The enhancement of the surface redshift is comparable to the magnitude of the anisotropy incorporated into the model. The role of anisotropy during dissipative collapse has yielded many interesting results (Maurya,  Banerjee \&  Hansraj 2018). Herrera and co-workers have shown that the anisotropy affects the dynamical instability of the star undergoing collapse (Herrera, Le Denmat \& Santos 1989; Chan, Herrera \& Santos 1993; Herrera, Le Denmat \& Santos 2012; Govender, Mewalal \& Hansraj 2019). The stability factor $\Gamma$ in both the Newtonian and post-Newtonian approximations deviate from the well-known result $(\Gamma > \frac{4}{3})$ first derived by Chandrasekhar (1964a). The sign of the anisotropic factor leads to a further deviation from the classical case. The presence of anisotropic stresses within the collapsing core can either advance or delay the formation of the horizon. From a thermodynamical point of view, it has been shown that pressure anisotropy leads to higher core temperatures. This effect is enhanced during the late stages of collapse when the differences in the anisotropic stresses are much larger (Govender et al. 2018).

 Note also that besides higher curvature effects induced by EGB quadratic invariants, other modifications of the standard theory are also vigorously studied. Amongst these is the $f(R, T)$ theory proposed by Harko {\it{et al}} (2011) where it is proposed that the action is dependent on the Ricci scalar $R$ as well as the trace of the energy-momentum tensor $T$. These ideas were further studied by Houndjo (2012), Jamil {\it{et al}} (2012), Chakraborty (2013) and Hansraj and Banerjee (2018) amongst others.  In addition Rastall (1972, 1976) advanced the idea that the divergence of the energy-momentum tensor could be proportional to the divergence of the Ricci scalar but in the process compromising the conservation of energy momentum as extrapolated from Newtonian physics. This was intensively investigated by Batista {\it{et al}} (2012, 2013), Fabris {\it {et al}} (2012), Silva {\it{et al}} (2013), Hansraj and Banerjee (2020) and Heydarzade {\it {et al}} (2017). A plausible explanation for the accelerated expansion of the universe problem came forth in the $f(R)$ theory of gravity with the proposition of Starobinksy (1980) of an action quadratic in the Ricci scalar.The drawback of this idea is that inevitably ghosts develop by virtue of the presence of derivatives of quartic order.  A comprehensive review of $f(R)$ theory may be found in Sotiriou and Faraoni (2010).

As pointed out above,  isotropic stellar models of perfect fluids suitable for modeling EGB  stars have recently been found. Such models, obtained on mathematical grounds,  lack a basic ingredient of astrophysical models namely an equation of state. It should be noted that even in the simpler case of Einstein's general relativity, no exact nontrivial isotropic models of stars involving an equation of state have been found analytically except for the over-determined isothermal model of Saslaw {\it{et al}} (1996).  Even in this case, the model has cosmological application since a surface of zero pressure is not admissible. All other explorations of isotropic models with an equation of state have been undertaken numerically, for example see Nilsson and Uggla (2001a, 2001b),  thereby introducing the prospects of approximation errors which are bound to be significant at the scale of stellar objects. Needless to say, equations of state when introduced into the more formidable EGB equations also lead to a mathematically intractable situation. On the other hand, to analyze the impact of equations of state, a relaxation in the condition of isotropy is required. The concomitant effect of sacrificing isotropy is that the system of equations become four in six unknowns - a seriously simpler mathematical problem. Even after prescribing an equation of state, there still remains a further choice to be made to close the equations. In this work we invoke a strange star equation of state. This avenue, for EGB stars,  does not appear to have been pursued  in the literature to date.  The equation of state contributes a genuine physical constraint on the model and augmenting this with the well studied superdense star potential ansatz, enhances the physical viability of the model.

Our intention in this paper is to solve the nonlinear EGB equations for a static spherically symmetric matter distribution with anisotropic stresses and with a strange star equation of state. Recently Panotopoulos and Rinc$\acute{o}$n (2019) studied isotropic deconfined quark star matter in the context of Einstein--Gauss--Bonnet theory. They concluded that the compactness of stars increases with the increase of the Gauss--Bonnet coupling parameter (of the order of unity) and suggested that this was significant in heavy stars.  The conclusions were based on a numerical integration of the TOV equation  and the standard four dimensional gravitational constant was used since the value of $G$ in higher dimensions and curvature is not known. The distinguishing feature of our analysis  is that we show how exact solutions of the field equations may be obtained for anisotropic matter with a generalized quark star equation of state. Exact solutions, unlike their numerical counterparts, do not suffer the constraints of the magnitude of approximations. Additionally, we study GB parameters of the order of $10^3$ and make conclusions on the very significant effects of the higher curvature terms on the surface gravitational redshift and the sound speed value.   In section II we briefly outline the basic equations in EGB gravity. The field equations in 5--dimensional EGB gravity are presented for a spherically symmetric metric, and they are then transformed to an equivalent form through a coordinate redefinition which helps in finding exact solutions. In Section III the generalized Vaidya--Tikekar (1982) superdense star ansatz is examined and a number of well known special cases are considered.  In Section IV the physical features of the Finch--Skea model are investigated with the help of graphical plots and a comparison with the 5 dimensional Einstein counterpart is made. We make use of data associated with the X--ray pulsar LMC X--4 in order to determine the values of constants in the problem and from the plots we deduce that the model displays the necessary qualitative features expected of such astrophysical objects.  Some concluding remarks are made in Section V.

\section*{Einstein--Gauss--Bonnet Gravity}

The   action
\begin{equation}
S=\frac{1}{2\kappa} \int R \sqrt{-g} d^4 x \label{1}
\end{equation}
is the standard Einstein--Hilbert action of general relativity. Here $g = \det (g_{ab}) $ is the determinant of the metric tensor $g_{ab}$, $R$ is the Ricci scalar and $\kappa = 8\pi Gc^{-4}$ where $G$ is the Newton's gravitational constant and $c$ is the speed of light in vacuum. The Lovelock tensor in $d$ dimensions may be written as
\begin{equation}
\mathcal{G}_{ab} = \sum ^{[(d-1)/2]}_{n=0} \alpha_n \mathcal{G}_{ab}^n  \label{5a}
\end{equation}
so that the Lovelock (1971, 1972) Lagrangian has the form
\begin{equation}
\mathcal{L} = \sum ^{n}_{n=0} \alpha_n \mathcal{R}^n  \label{5}
\end{equation}
where $ \mathcal{R}^n = \frac{1}{2^n} \delta^{c_1 d_1 ...c_n d_n}_{a_1 b_1 ... a_n b_n} \Pi^n_{r=1} R^{a_r b_r}_{c_r d_r} $ and $R^{a b}_{c d}$ is the Riemann or curvature tensor. Also $ \delta^{c_1 d_1 ...c_n d_n}_{a_1 b_1 ... a_n b_n} = \frac{1}{n!} \delta^{c_1}_{\left[a_1\right.} \delta^{d_1}_{b_1} ... \delta^{c_n}_{a_n} \delta^{d_n}_{\left.b_n \right]}$ is the required Kronecker delta. The quantity $[v]$ refers to the greatest integer value satisfying $[v] \leq v$. Note that $\mathcal{G}_{ab}$ is obtained by suitable contractions on a tensor product of $n$ copies of the Riemann tensor that trivially vanish whenever $n > [(d-1)/2]$. In the event that $d = 3, 4$, $\mathcal{G}_{ab}^n$ vanishes for all $n > 1$. The Lovelock terms become a total derivative or a topological invariant for $ d = 3, 4$ and hence do not contribute to the dynamics. Moreover, each term $\mathcal{R}^n$ in $\mathcal{L}$ represents the dimensional extension of the Euler density in $2n$ dimensions and contribute to the field equations only if $n < d/2$. For this reason, the critical spacetime  dimensions of Lovelock gravity are $d = 2n + 1$ and $d = 2n + 2$. In the case of EGB gravity $ n = 2$ so the critical dimensions are $d = 5, 6$. For $n = 3$, the salient dimensions are $7, 8$ and so on. A detailed treatment of this aspect may be found in the works of Beroiz {\it{et al}} (2007), Navarro and Navarro (2011) and Kastor (2013).

The quantity $ \alpha $ is the dimensionful Gauss--Bonnet coupling constant which may be identified with the string tension in string theory. Presently there are no experimental tests that constrain
the value of the coupling constant and moreover it is not known whether a should necessarily
be positive. In the work of Amendola {\it{et al}} (2007) it has been argued that the value of a
may be as high as of the order of $10^{23}$. These authors also considered solar system tests
in the context of Einstein--Gauss--Bonnet theory.   On the other hand, in order to explain the accelerated expansion of the universe, Dehghani (2004) used a negative coupling constant $\alpha$. Doneva and Yazadjiev \cite{doneva} have analyzed relativistic stars in the novel 4D EGB theory where the coupling constant undergoes a rescaling.  According to their numerical calculations values of $\alpha$ of unit order ($\alpha = -3, -2, -1, 0, 1, 2, 3$) are plausible. They also note that black holes with $\alpha < 0$ are stable and that the upper bound $\sqrt{|\alpha|} \leq 9.7$ km results.  Therefore according to these findings, values of $\alpha$ of unit order are reasonably speculated in the context of strong gravity regimes near black holes.  The strength of the action $ L_{G B} $ lies in the fact that despite the Lagrangian being quadratic in the Ricci tensor, Ricci scalar  and the Riemann tensor, the equations of motion turn out to be second order quasilinear which is consistent with a theory of gravity.

The Lovelock action (\ref{5}) may be expanded as
\[
\mathcal{L} = \sqrt{-g}\left( \alpha_0 + \alpha_1R + \alpha_2 \left( R^2 + R_{a b c d} R^{a b c d} - 4R_{c d} R^{c d} \right) + \alpha_3 \mathcal{O}(R^3)\right)
\]
in general. Up to second order of the Lovelock polynomial we  define the Gauss--Bonnet (GB) term as
\[
\mathcal{R}^2 = R^2 + R_{a b c d} R^{a b c d} - 4R_{c d} R^{c d}
\]
often denoted as $L_{GB}$.
This term  arises in the low energy effective action of heterotic string theory (Gross 1999).

The EGB field equations may be written as
\begin{equation}
G_{a b} + \alpha H_{a b} = T_{a b},  \label{2}
\end{equation}
where we have adopted the metric signature $ (- + + + +) $ and where $ G_{ab} $ is the usual Einstein tensor. The Lanczos tensor is given by
\begin{equation}
H_{a b} = 2 \left(R R_{a b} - 2 R_{a c}R^{c}_{b} - 2 R^{c d} R_{a c b d} + R^{c d e}_{a} R_{b c d e} \right) - \frac{1}{2} g_{a b} L_{G B}.  \label{3}
\end{equation}
Now varying the action against the metric generate the Einstein--Gauss--Bonnet equations of motion.

\section*{Field Equations}

Customarily the  five dimensional  metric for static spherically symmetric spacetimes is taken as
\begin{equation}
ds^{2} = -e^{2 \nu} dt^{2} + e^{2 \lambda} dr^{2} + r^{2} \left( d\theta^{2} + \sin^{2} \theta ~d \phi^2 + \sin^{2} \theta \sin^{2} \phi ~d\psi^2 \right), \label{5}
\end{equation}
where $ \nu(r) $ and $ \lambda(r) $ are  the gravitational potentials.
 The energy-momentum tensor for anisotropic matter may be expressed as
\begin{equation}
T_{ab} = (p_T+\rho)u_{a}u_b - p_T g_{ab}+(p_R - p_T)\chi_a\chi_b,\label{em}
\end{equation}
where $\rho$, $p_R$ and $p_T$ are the energy density, radial and transverse pressures respectively. We define the normalized 4-velocity vector  $u^{a}=\sqrt {\frac{-1}{g_{tt}}}\delta_{t}^{a}$ and the unit spacelike vector $\chi^a= \sqrt {\frac{1}{g_{rr}}}\delta_{r}^{a}$ along $r$ provided $g_{ab}u^{a}u^b = -1$ and $g_{ab}\chi^a\chi^b= 1$ respectively.
 The EGB field equations (\ref{2})  may then be written as the system
\begin{eqnarray}
\rho &=& \frac{3}{e^{4 \lambda} r^{3}} \left( 4 \alpha \lambda ^{\prime} +  r e^{2 \lambda} -  r e^{4 \lambda} -  r^{2} e^{2 \lambda} \lambda ^{\prime} - 4 \alpha e^{2 \lambda} \lambda ^{\prime} \right),  \label{6a} \\ \nonumber \\
p_R &=&  \frac{3}{e^{4 \lambda} r^{3}} \left(-  r e^{4 \lambda} + \left( r^{2} \nu^{\prime} +  r + 4 \alpha \nu^{\prime} \right) e^{2 \lambda} - 3 \alpha \nu^{\prime} \right),  \label{6b} \\ \nonumber \\
p_T &=& \frac{1}{e^{4 \lambda} r^{2}} \left( -e^{4 \lambda} - 4 \alpha \nu^{\prime \prime} + 12 \alpha \nu^{\prime} \lambda^{\prime} - 4 \alpha \left( \nu^{\prime} \right)^{2}  \right)  + \frac{1}{e^{2 \lambda} r^{2}} \left(  1 - r^{2} \nu^{\prime} \lambda^{\prime} + 2 r \nu^{\prime} - 2 r \lambda^{\prime} + r^{2} \left( \nu^{\prime} \right)^{2}  \right) \nonumber \\
                 & \quad & + \frac{1}{e^{2 \lambda} r^{2}} \left(  r^{2} \nu^{\prime \prime} - 4 \alpha \nu^{\prime} \lambda^{\prime} + 4 \alpha \left( \nu^{\prime} \right) ^{2} + 4 \alpha \nu^{\prime \prime}   \right). \label{6c}
\end{eqnarray}
where the subscripts $R$ and $T$ refer to the radial and transverse components respectively.
  The equations (\ref{6a})--(\ref{6c}) constitute a system of  three differential  equations in five variables namely, the density, radial pressure, tangential pressure and two gravitational potentials $\nu$ and $\lambda$. In this form it is easy to see that any arbitrary metric solves the system which is under-determined. This approach is however unlikely to yield  exact models that conform to the elementary tests for physical viability. Accordingly inserting some constraints of physical importance will likely give solutions that may be used to model compact stars. In this work, we prescribe a strange star equation of state and this immediately increases the mathematical complexity. However, there remains one more prescription to make to close the system. We shall employ a metric ansatz of a superdense star in order to determine a unique solution.  Observe that the  vacuum metric describing the gravitational field exterior to the 5--dimensional static perfect fluid may be described by the Boulware--Deser (1985) spacetime as
\begin{equation}
ds^2 = - F(r) dt^2 + \frac{dr^2}{F(r)} + r^{2} \left( d\theta^{2} + \sin^{2} \theta ~d \phi^2 + \sin^{2} \theta \sin^{2} \phi ~d\psi^2 \right), \label{7}
\end{equation}
where
\[
F(r) = 1 + \frac{r^2}{4\alpha} \left( 1 - \sqrt{1 + \frac{8m\alpha}{r^4}} \right).
\]
In the above $ m $ is associated with the gravitational mass of the hypersphere. The exterior solution is  unique up to branch cuts, however,  there appears to be no equivalent of the Birkhoff theorem of the 4--dimensional Einstein gravity case. Bogdanos et al. (2009) have investigated the 6--dimensional case in EGB and demonstrated that Birkhoff's theorem holds for particular assumptions. At this point we also note that the Buchdahl (1959) compactness limit  for a perfect fluid sphere $\frac{M}{R} = \frac{4}{9}$  was recently improved to the case of 5 dimensional EGB (Wright 2016) but the results depend on the sign of the coupling constant $\alpha$.

To enhance our chances of locating exact solutions, we make the following change of variables  $ e^{2 \nu} = y^{2}(x) $, $ e^{-2 \lambda} = Z(x)  $ and $ x = C r^{2} $ ($ C $ being an arbitrary constant). This set of transformations has proved particularly useful in the case of isotropic fluids since the isotropy equation may be written as linear differential equations in either variable $y$ or $Z$ in Einstein gravity. In EGB, the same equation is linear in $y$ but nonlinear in $Z$.   For applications of this approach  to charged anisotropic relativistic matter see the recent works of Mafa Takisa \& Maharaj (2013) and Maharaj, Sunzu \& Ray (2014) in four dimensional Einstein theory. The field equations (\ref{6a})--(\ref{6c}) may now  be expressed  as
\begin{eqnarray}
-3  \dot{Z} - \frac{3  (Z - 1) ( 1 -  \beta \dot{Z} )}{x} &=& \frac{\rho}{C}, \label{8a} \\ \nonumber \\
\frac{3  (Z - 1)}{x} + \frac{6  Z \dot{y}}{y} - \frac{6\beta (Z - 1) Z \dot{y}}{x y} &=& \frac{p_R}{C}, \label{8b} \\ \nonumber \\
4 Z \left[ \beta (1-Z) +x \right] \frac{\ddot{y}}{y}
+ \left[ \frac{2\beta Z(1-Z)}{x} + 2(x +\beta)\dot{Z} +6Z(1-\beta \dot{Z} \right] \frac{\dot{y}}{y}  +\left[\frac{Z-1}{x} + 2\dot{Z} \right]   &=& \frac{p_T}{C},   \label{8c}
\end{eqnarray}
where we have introduced the constant $\beta = 4\alpha C$ containing the EGB coupling constant.

We now utilise a physically important equation of state relating the density and pressure. The prescription $p_R = \gamma\rho - \xi$ is understood to be valid for strange star material or quark stars  which have a higher density and larger rotation than neutron stars. The special case $\gamma = \frac{1}{3}$ corresponds to the well studied MIT Bag model in four dimensions where quarks are considered as free particles and their thermodynamic properties are generated by treating them as an  Fermi (ideal) gas. Panotopoulos and Rinc$\acute{o}$n (2019) show from standard thermodynamics that for five dimensional spacetime, the de-confined quark star matter with the applicable MIT bag model equation of state modifies as $p = \frac{1}{4}(\rho - 5B)$ where $B$ is the bag constant. We have elected to retain the bag coefficient as $\frac{1}{3}$ and introduced a generic constant $\xi$ for strange matter not necessarily representing de-confined quark matter.   With this equation of state (\ref{8a}) and (\ref{8b}) together   yield
\be \frac{\dot{y}}{y} = \left[6Z - \frac{6\beta(Z-1)Z}{x}\right]^{-1}\left[3\gamma {\dot Z} + \frac{3(Z - 1)(\gamma - 1 - \beta\gamma Z)}{x} - \xi\right] \label{10}
\ee
 where $\gamma$ and $\beta \geq 0$ are constants. Equation (\ref{10}) integrates as
 \be
 y= C_1\exp \left(\int \frac{3\gamma x{\dot Z} + 3(Z - 1)(\gamma - 1 - \beta\gamma Z) - \xi x}{6xZ - 6\beta (Z-1)Z} dx \right)  \label{11}
 \ee
where $C_1$ is an integration constant. It now remains to detect forms for $Z$ that will permit the complete integration of (\ref{11}).

\section*{Vaidya-Tikekar superdense star ansatz}

Equation (\ref{11}) admits a large number of potentials $Z$ for which an exact solution exists. Therefore, it is prudent to make a selection from well studied models which are known to be physically reliable. Expressed in terms of our coordinates the generalized Vaidya--Tikekar potential prescription, known to generate super-dense stellar models (Vaidya \& Tikekar 1982),  is given by
\be
Z = \frac{1+ax}{1+bx}  \label{12}
\ee
where $a$ and $b$ are arbitrary real numbers related to the spheroidal parameter. The Vaidya--Tikekar spatial potential ansatz has been used by various researchers in different contexts and the reader may consult \cite{vt1,vt2,vt3,vt4} and the references therein.   Note that specifying the spatial metric potential is tantamount to determining the law of variation of the density profile. The special case b = 0 corresponds to the constant density Schwarzschild interior solution in both Einstein and Einstein-Gauss-Bonnet gravity and is moreover independent of spacetime dimension. These results were first reported by Dadhich (2010) and more thoroughly discussed by Hansraj {\it {et al}} (2021) who obtained all static conformally flat solutions in EGB.  It turned out from this analysis that the traditional Schwarzschild interior metric was one of two possible solutions for conformal flatness in EGB whereas it is the unique conformally flat solution in Einstein gravity. In both cases the incompressible Schwarzschild metric arises from the potential $Z = 1 + ax$.    The case a = 0 is the Finch{Skea (1989) ansatz first proposed by Duorah and Ray (1987). An exact solution for spheroidally distributed matter was examined in the case $a = -1$ and $b = 2$ by Vaidya and Tikekar (Vaidya \& Tikekar 1982) and shown to admit models with surface densities $2 \times 10^{14}$ $g/cm^3$ with masses of about 4 times the solar mass. The choice $b =1$ was studied by Buchdahl (1959,1984) and recently Molina {\it {et al}} used this ansatz to find models of stars in pure Gauss--Bonnet gravity (Molina, Dadhich \& Khugaev 2017).    The general integral of (\ref{11}) has the form
\beq
y &=& C_1 \exp \left( \frac{1}{6} \left(a^2\left(\left(-3 a^2 \beta ^2 \gamma +a \beta  (b (3 (\beta +1) \gamma -\beta  \xi -3)+3 \gamma )    \right. \right. \right. \right. \n \\ \n \\ && \left. \left. \left. \left.   +     b (b \beta  (\beta  \xi -3 \gamma +3)
 - 3 (\beta -1) \gamma )\right) \log (-a \beta +b (\beta +x)+1)\right)  \right. \right. \n \\ \n \\  && \left. \left.
-b\left(\log (a x+1) \left(a^2 (6 \gamma -3)-a (3 b (\gamma -1)+\xi )+b \xi \right)\right)-ab^2 \xi  x\right) /(ab(a\beta -1)) \right) \label{13}
\eeq
for the potential (\ref{12}).

\subsection*{Finch Skea spatial potential}

 The Finch--Skea potential $Z =  \frac{1}{1+x} = \frac{1}{1+Cr^2}$  was used to model four dimensional static stars with behaviors consistent with the astrophysical theory of Walecka (1975). It is also well known  that for regular stars, that is models that are singularity-free, it is necessary that the spatial potential has the form $ 1 + O(r^2)$. This proves to be useful in this higher curvature analysis as well and it will be observed that all physical quantities are free of the defect of being singular somewhere within the distribution.  For the Finch--Skea prescription  the potential (\ref{13}) assumes the simplified form
\beq
y &=& C_1 (\beta  +w_1)^{a_1} e^{-\frac{1}{12} \left(a_2+\xi  x\right) (\beta  + w_1)}
\eeq
where we make the substitutions $a_1 =\frac{1}{6} (3 \beta  \gamma -\beta   (\beta   \xi -3 \gamma +3)-3 \gamma )$, $a_2 = 6\xi\gamma -3\xi\beta + \xi -6$, $w_1 = 1+x $, $w_2 = 1+x+ \beta$ and $w_3 = 1+2x+\beta$ to shorten the lengthy expressions to follow.
The associated dynamical quantities have the form
\beq
\frac{\rho}{C} &=& \frac{3 (\beta +x (w_1+2)+2)}{w_1^3} \label{15a}\\ \n \\
\frac{p_R}{C} &=&  \frac{12 a_1-a_2 w_2-\xi  w_2 w_3-6 w_1}{2 w_1^2}  \label{15b} \\\n\\
\frac{p_{T}}{C} &=& \left[ -6 \xi  \left(3 (\beta +1)^2+8 x^3+(6 \beta +20) x^2+15 (\beta +1) x\right)-36 w_1 (x+3) +2 \xi  x w_1 w_2 w_3 \right. \n\\
&& \left. +\frac{144 a_1^2 x w_1}{w_2}\xi ^2 +x w_1 w_2 w_3^2  +a_2 \left(a_2 x w_1 w_2-6 \left(3 \beta +2 x^2+5 x+3\right)\right) \right] \mathbin{/} 36 w_1^3
 \label{15c}
\eeq
while the measure of the pressure anisotropy $\Delta = p_T - p_{R}$ is given by the expression
\beq
\frac{\Delta}{C} &=&  \left[72 w_1-6 \xi  \left(-3 \beta ^2-3 \beta  x+x (2 x+5)+3\right)+\xi ^2 w_1 w_2 w_3^2 \right. \n\\
&& +\frac{144 a_1^2 w_1}{w_2}-24 a_1 \left(a_2 w_1+\xi  w_1 w_3+9\right) \n\\
&& \left. +a_2 \left(a_2 w_1 w_2+2 \xi  w_1 w_2 w_3+6 (3 \beta +x+1)\right) \right] \mathbin{/} 36 w_1^2   \label{15e}
\eeq
Observe that a hypersurface of vanishing pressure exists when $p_R = 0$ demarcating the boundary of the 5 dimensional hypersphere at
\be
x=\frac{1-2 a_1 \beta -2 a_1-2 a_2}{2 a_1-1}
\ee
in terms of the constants associated with the strange star equation of state and Gauss--Bonnet coupling constant.
The ratio of the pressure to the energy density  $ \frac{p}{\rho}$ is  understood to give an indication of the equation of state of the model. In this case we obtain
\beq
{\left(\frac{p}{\rho}\right)}_R &=& \frac{w_1 \left(12 a_1-a_2 w_2-\xi  w_2 w_3-6 w_1\right) }{6 (\beta +x (x+3)+2)}   \n\\ \label{15f} \\
{\left(\frac{p}{\rho}\right)}_{T}&=& \left[ -36 w_1 (x+3)-6 \xi  \left(3 (\beta +1)^2+8 x^3+(6 \beta +20) x^2+15 (\beta +1) x\right) \right. \n\\
&& \left. +\frac{144 a_1^2 x w_1}{\beta +x+1}+\xi ^2 +x w_1 w_2 w_3^2  -24 a_1 \left(a_2 x w_1+\xi  x w_1 w_3-9\right) \right. \n\\
&& + \left. a_2 \left(a_2 x w_1 w_2-6 \left(3 \beta +2 x^2+5 x+3\right)+2 \xi  x w_1 w_2 w_3\right) \right] \mathbin{/} 108 (\beta +x (x+3)+2) \n\\ \label{15g}
\eeq
for the radial and transverse components.

\section*{Viability Tests}

In what follows, we analyze a variety of tests usually imposed on stellar models to test their physical applicability.

\subsection*{Causality}

The causal behavior of stars is studied by examining the square of the sound speed given by the formula  $v^2 = \frac{dp}{d\rho}$. This evaluates to
\beq
v^2_R &=& -\frac{w_1 \left(a_2 (2 \beta +x+1)-24 a_1+\xi  \left(2 \beta ^2+\beta +3 \beta  x-x-1\right)+6 w_1\right)}{6 \left(3 \beta +x^2+4 x+3\right)} \label{15h} \\ \n\\
v^2_{T}  &=& - \{ \left.36 w_1 (x+5)+6 \xi  \left(9 \beta ^2+3 \beta +6 \beta  x (x+3)-2 w_1 (2 x+3)\right) \right.\n\\
&& +\xi ^2w_1 w_3 \left((\beta +1)^2+4 x^3+2 (\beta +5) x^2-(\beta -7) (\beta +1) x\right) \n\\
&& -\frac{144 a_1^2 w_1 \left(2 x^2+\beta  (x-1)+x-1\right)}{w_2^2}\n\\
&& +24 a_1 \left(a_2 \left(x^2-1\right)+\xi  w_1 (\beta  (x-1)-3 x-1)-27\right) \n\\
&& +a_2\left[2 \xi  w_1 \left((\beta +1)^2+2 x^3+6 x^2-(\beta -5) (\beta +1) x\right)+6 (9 \beta +2 x (x+3)+4) \right] \n\\
&& +a_2^2 (\beta +x (-\beta  x+x+2)+1) \} \mathbin{/}  108 \left(3 \beta +x^2+4 x+3\right) \n\\  \label{15i}
\eeq
and the expectation is that both these quantities should be constrained in the interval $(0; 1)$ to guarantee that the sound speed remains subluminal. The possibility of superluminal behavior  in ultrabaric matter in special relativity was discussed by Caporaso and Bescher (1979) and ruled out.

\subsection*{Cracking phenomenon}

Herrera (1992) established the cracking concept applicable to local anisotropic stars. According to this principle anisotropic distributions are stable provided that $0 \leq |v_t^2 - v_R^2| \leq 1$ is satisfied.
The difference between the squares of the radial and transverse sound speeds given by
\beq
v^2_R - v^2_{T} &=& \{ 324 w_1^2 \left[a_2 (2 \beta +x+1)-24 a_1+\xi  \left(2 \beta ^2+\beta +3 \beta  x-x-1\right)+6 w_1\right]{}^2 \n\\
&& - \left[ 6 \xi  \left(9 \beta ^2+3 \beta +6 \beta  x (x+3)-2 w_1 (2 x+3)\right)+36 w_1 (x+5) \right. \n\\
&& +\xi ^2w_1 w_3 \left((\beta +1)^2+4 x^3+2 (\beta +5) x^2-(\beta -7) (\beta +1) x\right) \n\\
&& -\frac{144 a_1^2 w_1 \left(2 x^2+\beta  (x-1)+x-1\right)}{w_2^2} \n\\
&& +24 a_1 \left(a_2 \left(x^2-1\right)+\xi  w_1 (\beta  (x-1)-3 x-1)-27\right) \n\\
&& + a_2\left(2 \xi  w_1 \left((\beta +1)^2+2 x^3+6 x^2-(\beta -5) (\beta +1) x\right)+6 (9 \beta +2 x (x+3)+4)\right)\n\\
&& \left. a_2^2(\beta +x (-\beta  x+x+2)+1) \right]^2\} \mathbin{/} 11664 \left(3 \beta +x^2+4 x+3\right)^2 \label{15y}
\eeq
provides an indication of the stability of the model. Graphical plots will be used to analyze these features.

\subsection*{Compactification ratio}

The active gravitational mass is computed via the formula $ \frac{1}{3}\int \rho r^{d-2} dr$ where $d$ is the spacetime dimension. In the five dimensional case we obtain
\beq
M(r) &=& \frac{k}{3} + \frac{1}{2C^2}\left( x-\frac{\beta +2 (\beta -1) x-2}{2 w_1^2} \right) \label{15j}
\eeq
and correspondingly the compactification parameter
\beq
\frac{M(r)}{r} &=& \left[ \frac{k}{3} + \frac{1}{2C^2}\left(  x-\frac{\beta +2 (\beta -1) x-2}{2 w_1^2}\right)  \right] \times \sqrt{\frac{c}{x}}  \label{15k}
\eeq
will be useful in determining whether the Buchdahl limit for the mass-radius ratio applicable to Einstein stars is still valid when higher curvature effects are included.

\subsection*{Adiabatic stability of Chandrasekhar}

Another indicator of stability devised by Chandrasekhar (1964a, 1964b) is the adiabatic stability parameter $\Gamma = \left(\frac{\rho + p}{p}\right) \frac{dp}{d\rho}$ which assume the forms
\beq
\Gamma_R&=&  \frac{\left(a_2 (2 \beta +x+1)-24 a_1+\xi  \left(2 \beta ^2+\beta +3 \beta  x-x-1\right)+6 w_1\right) }{6 \left(3 \beta +x^2+4 x+3\right) \left(a_2 w_2-12 a_1+\xi  w_2 w_3+6 w_1\right)} \n\\
&& \times \frac{\left(w_2 \left(a_2 w_1+\xi  w_1 w_3-6\right)-12 a_1 w_1\right)}{6 \left(3 \beta +x^2+4 x+3\right) \left(a_2 w_2-12 a_1+\xi  w_2 w_3+6 w_1\right)}  \label{15l} \\\n\\
\Gamma_{T} &=& -\{ 36 \left(3 \beta +2 x^2+5 x+3\right)-6 \xi  \left(3 (\beta +1)^2+8 x^3+(6 \beta +20) x^2+15 (\beta +1) x\right)  \n\\
&& -24 a_1 \left(a_2 x w_1+\xi  x w_1 w_3-9\right)+\frac{144 a_1^2 x w_1}{\beta +x+1}+\xi ^2 +x w_1 w_2 w_3^2 \n\\
&&  +a_2 \left(a_2 x w_1 w_2-6 \left(3 \beta +2 x^2+5 x+3\right)+2 \xi  x w_1 w_2 w_3\right) \} \n\\
&& \times \{ 6 \xi  \left(9 \beta ^2+3 \beta +6 \beta  x (x+3)-2 w_1 (2 x+3)\right)+36 w_1 (x+5)  \n\\
&& + \xi ^2 w_1 w_3 \left((\beta +1)^2+4 x^3+2 (\beta +5) x^2-(\beta -7) (\beta +1) x\right) \n\\
&& +24 a_1 \left(a_2 \left(x^2-1\right)+\xi  w_1 (\beta  (x-1)-3 x-1)-27\right) \n\\
&& -\frac{144 a_1^2 w_1 \left(2 x^2+\beta  (x-1)+x-1\right)}{w_2^2} \n\\
&& +a_2\left[2 \xi  w_1 \left((\beta +1)^2+2 x^3+6 x^2-(\beta -5) (\beta +1) x\right)+6 (9 \beta +2 x (x+3)+4) \right] \n\\
&&  +a_2^2 (\beta +x (-\beta  x+x+2)+1) \} \mathbin{/} \{ 108 (3 \beta +x^2+4 x+3) \n\\
&& \times  \left[ -6 \xi  \left(3 (\beta +1)^2+8 x^3+(6 \beta +20) x^2+15 (\beta +1) x\right)-36 w_1 (x+3) \right. \n\\
&& +\frac{144 a_1^2 x w_1}{\beta +x+1}+\xi ^2 +x w_1 w_2 w_3^2  -24 a_1 \left(a_2 x w_1+\xi  x w_1 w_3-9\right) \n\\
&& \left. +a_2\left(2 \xi  x w_1 w_2 w_3-6 \left(3 \beta +2 x^2+5 x+3\right)\right)  +a_2^2 x w_1 w_2 \right] \} \n\\ \label{15x}
\eeq
for the anisotropic model under consideration. Adiabatic stability occurs provided that  $\Gamma$ exceeds the critical value $\frac{4}{3}$. For a recent study of this property  in the context of neutron stars see the work of Koliogiannis \& Moustakidis (2019).

\subsection*{Gravitational redshift}

The gravitational surface redshift $z$ obtained from the formula $z = e^{-\nu} -1$ is given by
\beq
z &=& \frac{1}{A} w_2^{-a_1} e^{\frac{1}{12} w_2 \left(a_2+\xi  x\right)} -1   \label{15m}
\eeq
for our model.

\subsection*{Energy conditions}

The energy conditions for anisotropic matter may be investigated with the help of the expressions $\rho - p$ (weak energy condition, $\rho + p$ (strong energy condition and $\rho + 3p$ the dominant energy condition.  For the radial and transverse directions we obtain
\beq
\frac{\rho - p_R}{C} &=& { w_1 \left(a_2 w_2-12 a_1+\xi  w_2 w_3+6 w_1\right)+6 (\beta +x (x+3)+2) \over 2 w_1^3} \label{15o} \n\\\\
\frac{\rho - p_{T}}{C}  &=&  \{ 36 \left(3 \beta +4 x^2+13 x+9\right)+6 \xi  \left[3 (\beta +1)^2+8 x^3+(6 \beta +20) x^2+15 (\beta +1) x\right] \n\\
&&  -\xi ^2 x w_1 w_2 w_3^2-\frac{144 a_1^2 x w_1}{\beta +x+1} +24 a_1 \left(a_2 x w_1+\xi  x w_1 w_3-9\right) - a_2^2 x w_1 w_2 \n\\
&& +a_2\left[18 (\beta +1)-2 \xi  x w_1 w_2 w_3+6 x (2 x+5)\right]\} \mathbin{/} 36 w_1^3 \label{15p} \\ \n\\
\frac{\rho + p_R}{C} &=& \frac{12 a_1 w_1-w_2 \left(a_2 w_1+\xi  w_1 w_3-6\right)}{2 w_1^3} \label{15q} \n\\\\
\frac{\rho + p_{T}}{C} &=& \{ 36 \left(3 \beta +2 x^2+5 x+3\right)-6 \xi  \left[3 (\beta +1)^2+8 x^3+(6 \beta +20) x^2+15 (\beta +1) x\right]\n\\
&& +\frac{144 a_1^2 x w_1}{\beta +x+1}+\xi ^2 +x w_1 w_2 w_3^2 -24 a_1 \left(a_2 x w_1+\xi  x w_1 w_3-9\right) \n\\
&& +a_2 \left[a_2 x w_1 w_2-6 \left(3 \beta +2 x^2+5 x+3\right)+2 \xi  x w_1 w_2 w_3\right]\} \mathbin{/} 36 w_1^3 \label{15w} \\  \n\\
\frac{\rho + p_R+3p_T}{C} &=&  \{ 3 \xi  \left[9 (\beta +1)^2+22 x^3+(21 \beta +55) x^2+3 (\beta +1) (\beta +14) x\right]+18 \left[3 (\beta +5) \right. \n\\
&& \left. +x (8 x+23)\right] -\xi ^2 x w_1 w_2 w_3^2-\frac{144 a_1^2 x w_1}{\beta +x+1} +12 a_1 \left(2 a_2 x w_1+2 \xi  x w_1 w_3-9 (x+3)\right) \n\\
&&  +a_2 \left[-a_2 w_1 x w_2+21 x^2-2 \xi  w_1 x w_2 w_3+9 \beta  (x+3)+48 x+27\right] \} \mathbin{/} 18 w_1^3 \label{15r} \n\\
\eeq

\subsection*{Matching}

In order to study the matching problem in EGB, it is necessary that the first and second fundamental forms are continuous across a common boundary hypersurface. The general approach was discussed by Davis (2003) however to date no applicable scheme to implement has come forth for EGB gravity. The matching conditions generate a set of intractable field equations. In contrast, the problem was adequately dealt with in general relativity by Israel (1966) and Darmois (1927) who went on to show that the matching of the second fundamental forms is tantamount to the vanishing of the radial pressure. In the area of $f(R)$ theory, a similar analysis by Goswami {\it{et al}} generated a set of five conditions that must be satisfied for a successful matching. The same authors opined that these strong constraints  render models of collapsing stars in $f(R)$ gravity unphysical. Additionally Senovilla (2013) considered the junction conditions relating to matter shells or branes in $f(R)$ theory.  In this same context De la Cruz-Dombriz {\it{et al}}  (2014) derive a set of junction conditions by requiring that thin shells do not form in the bulk equations of motion.  Resco {\it {et al}} examined the matching problem and utilised a fourth order Runge-Kutta scheme for a perturbative analysis. In order to execute the algorithm it was assumed that the pressure vanishes on the boundary and that the usual requirements of general relativity, namely Minkowskian flatness at spatial infinity were demanded.   These results differ from general relativity in general and if continuity of scalar torsion is satisfied then the GR condition is regained. In the case of $f(T)$ gravity, where $T$ refers to torsion,  Velay-Vitow and deBenedictis (2017) derive a set of junction conditions from the variational principle in the covariant version of the $f(T)$ theory.

 In our case we match the interior spacetime with the exterior geometry. Note that there are only two free constants to fix namely $C$ and $C_1$. The matching entails writing all integration constants in terms of the mass $M$ and radius  $R$ of the star. For this purpose we set $r = R$ and $m = M$  in (\ref{7}) and where capital letters now denote the values of the variables at the boundary interface. The matching of the $g_{00}$ component across the boundary allows us to express the constant $C$ in the form
\be
C = \frac{\sqrt{R^4 + 8\alpha M} -R^2}{4\alpha + R^2 - \sqrt{r^4 + 8\alpha M}} \label{161a}
\ee
while the matching of the $g_{11}$ component gives
\be
C_1 = \frac{\sqrt{4\alpha + R^2 - \sqrt{R^4 + 8\alpha M}} \exp\left( {\frac{1}{12} (a_2 + \xi CR^2)(\beta + 1 + CR^2)}\right)}{2\sqrt{\alpha} (\beta + 1 + CR^2)^{a_1}} \label{161b}
\ee
as the form of the integration constant in terms of $M$ and $R$.  The matching is now complete. Observe that the matching of the metric potentials is only the minimum criterion for continuity across the boundary hypersurface. Additional constraints may arise from the Davis (2003) conditions once these are explicitly known for spherically symmetric stars in EGB.

\section*{Physical Analysis}

In this section we discuss the physical plausibility of our compact star model. In order to generate the plots we have utilized mass and radius data associated with the pulsar LMC X-4 which qualifies as a superdense star to determine integration constants while other constants were assigned special values through fine--tuning. Specifically we have utilized values of the Gauss--Bonnet coupling $\alpha$ of the order of $10^{3}$ as these values generate physically pleasing plots. Note that in this investigation we are using geometric units in which the gravitational constant $G$ and the speed of light $c$ are both set to unity. For this reason we cannot make quantitative determinations on the value of $\alpha$. The study does offer us an avenue to make qualitative conclusions from the illustrative values of $\alpha$ and to contrast with the Einstein scenario when $\alpha = 0$. Additionally it should be noted that it is justifiable to use the data from observed 4 dimensional objects in the five dimensional scenario because the additional dimensions are angular and historically extra dimensions have been understood to be topologically hidden having very small values. This is the case for Kaluza--Klein theory and brane-world scenarios.

 Let us proceed with the analysis of  the plots of the physically relevant quantities.   In Fig. 1 the density is shown to be a smooth singularity--free monotonically decreasing function of the radial coordinate. We observe that the density decreases with an increase in magnitude of the coupling constant. Figure 2 shows that the radial and tangential pressures decrease monotonically outwards towards the stellar surface. This is expected as the density in the central regions of the star is much higher than the surface density. It is interesting to note that the radial pressure is greater than the tangential pressure for large values of the coupling constant. This means that the force due to the pressure anisotropy is attractive in this regime. As the coupling constant decreases the tangential pressure dominates the radial pressure leading to a repulsive contribution from the anisotropy. Most importantly a hypersurface of vanishing radial pressure is clearly visible for a radial value of approximately 8.3 km. The behavior of all physical quantities should be studied within this radius. The central pressure is well behaved displaying no singularities for any value of the coupling constant $\alpha$.   Observations of the adiabatic stability index $\Gamma$ in Figure 3 show that the fluid is more stable for increasing $\alpha$. This implies that higher order corrections tend to make the compact object more stable against perturbations. The critical lower bound of $\frac{4}{3}$ established by Chandrasekhar for Einstein gravity is always exceeded for both tangential and transverse directions.  Causality is obeyed throughout the fluid configuration as exhibited in Figure 4. Both the radial and tangential speeds of sound lie within the bound $(0, 1)$. The equation of state parameter is an important indicator of the relationship between the pressure and density at each interior point of the star. From Figure 5 we observe that the ratio of the pressure to density increases with an increase in the coupling constant. This implies that  stronger contributions from higher order corrections lead to more compact objects. In Figure 6 we observe that the anisotropy changes sign which implies that the force associated with anisotropy can be repulsive ($p_T > p-R$) or attractive ($p_T < p_R$). Lower order contributions (smaller values of $\alpha$) lead to repulsive effects due to anisotropy. All the energy conditions  as displayed in Figure 7 are satisfied. The metric potentials are continuous and well-behaved throughout the star as evidenced in Figure 8. The surface redshift is illustrated in Figure 9. We observe that the surface redshift is higher for smaller values of $\alpha$ which supports our observation of the density increasing with smaller values of the coupling constant (Figure 1). Figure 10 depicts the  cracking stability condition which is well behaved.   The forces required for equilibrium are illustrated in Figure 11. In order to achieve equilibrium we require that $F_g + F_h + F_a = 0$ where $F_g$, $F_h$ and $F_a$ are the gravitational, hydrostatic and anisotropic forces respectively. Figure 12 displays the variation of the mass with respect to the radius. Clearly within the stellar radius 8.3 km there appears to be little difference in the mass profile for various $\alpha $ values. If the gravitational field admitted a higher radial value then it can be inferred from  the plot  that a maximum mass is achieved and some discrimination in values  near this maximum occur.

The frames in Figure 13 to Figure 15 display the various physical quantities of our compact model in the 5D classical Einstein limit (ie., $\alpha = 0$). Radial quantities are in red while transverse items are in black.  We observe that the respective quantities such as density, pressures, redshift and anisotropy are all substantially higher than their EGB counterparts hence the need for separate plots. Xian--Feng \& Huan--Yu (2014) established  through relativistic mean field theory that the surface gravitational redshift of the star PSR J0348+0432 is in the region of about 0.3473 to 0.4064 which was higher than the canonical mass neutron star with a redshift of 0.226. Our 5D stellar model (Figure 14) displays a redshift in the range 0.15 to 0.22 within the distribution. This is therefore comparable with a neutron star. Note that when higher curvature terms are present as depicted in Figure 9, the surface redshift drops dramatically to the range of order 0.0025 to 0.0050. Figure 14 also demonstrates that the measure of anisotropy, the energy conditions and the speed of sound are all within the expected levels. The stability measures shown in Figure 15 confirm that the 5D Einstein model is stable with a well behaved mass profile. These stability features are not disturbed by the introduction of higher curvature effects due to the Gauss--Bonnet action. In Table 1 we exhibit a few stellar models which fall in the range of the mass and radius comparable to LMC X--4. This shows that our qualitative results are consistent with a large number of known stars.

\begin{figure}[h]
	\centering
	\includegraphics[width=0.5\linewidth]{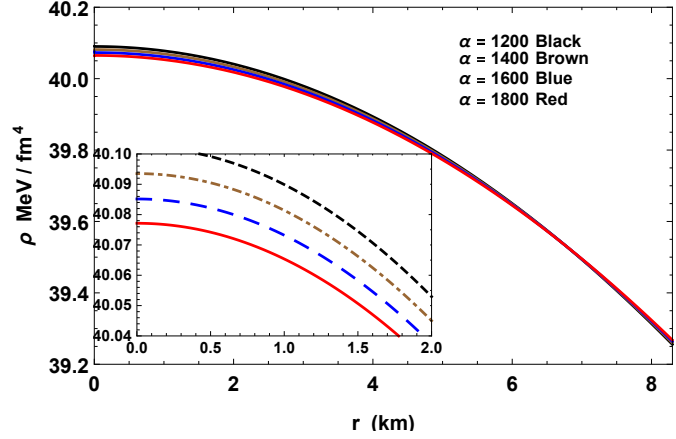}
	\caption[Density]{Variation of density $\rho$ with radial coordinate for LMC X-4 with $M = 1.04 M_{\odot}, ~R = 8.3 km$ and $\gamma = 1/3$ in EGB.}
	\label{fig:den}
\end{figure}
\begin{figure}[h]
	\centering
	\includegraphics[width=0.5\linewidth]{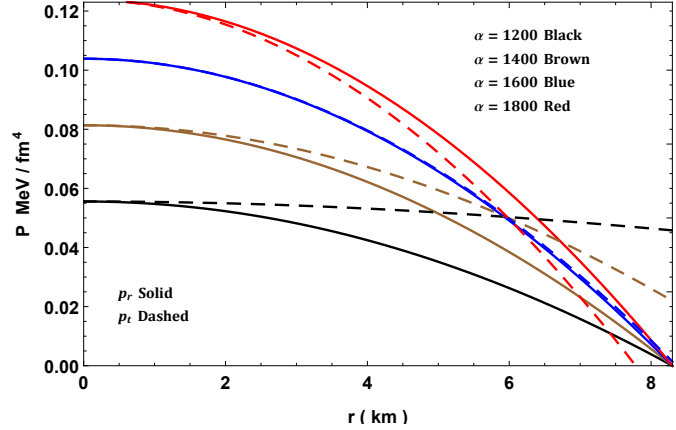}
	\caption[Pressure]{Variation of pressures $p$ with radial coordinate for LMC X-4 with $M = 1.04 M_{\odot}, ~R = 8.3 km$ and $\gamma = 1/3$ in EGB.}
	\label{fig:pre}
\end{figure}
\begin{figure}[h]
	\centering
	\includegraphics[width=0.5\linewidth]{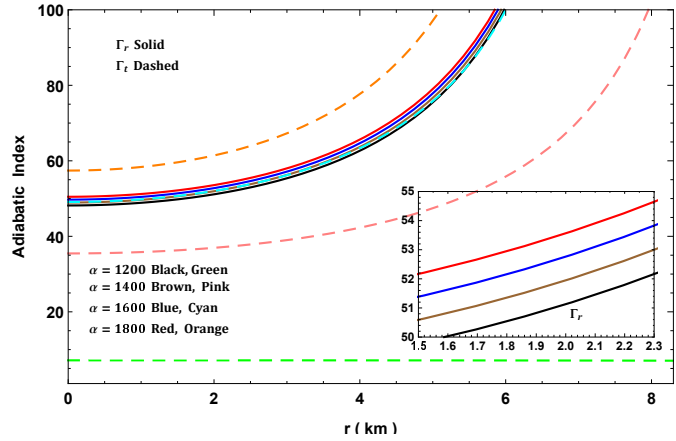}
	\caption[Adiabatic Index]{Variation of adiabatic index $\Gamma$ with radial coordinate for LMC X-4 with $M = 1.04 M_{\odot}, ~R = 8.3 km$ and $\gamma = 1/3$ in EGB.}
	\label{fig:gam}
\end{figure}
\begin{figure}[h]*
	\centering
	\includegraphics[width=0.5\linewidth]{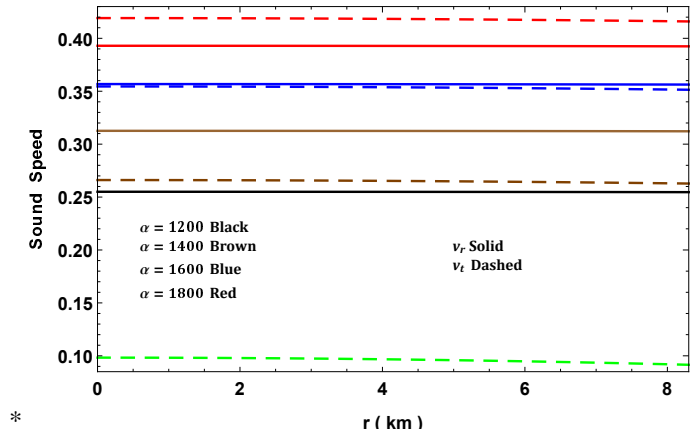}
	\caption[Sound Speed]{Variation of sound speed with radial coordinate for LMC X-4 with $M = 1.04 M_{\odot}, ~R = 8.3 km$ and $\gamma = 1/3$ in EGB.}
	\label{fig:sou}
\end{figure}
\begin{figure}[h]
	\centering
	\includegraphics[width=0.5\linewidth]{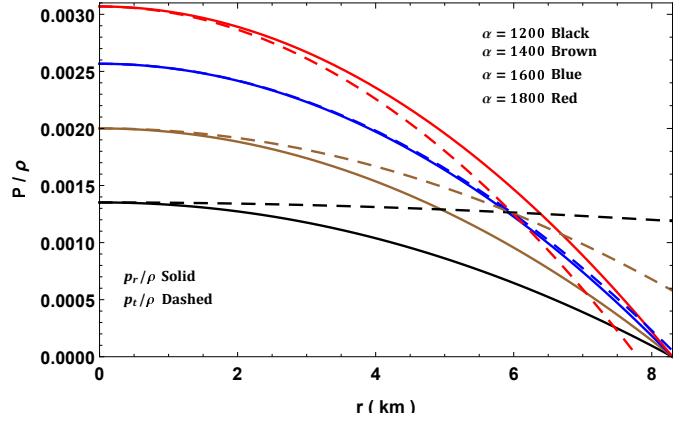}
	\caption[Equation of State]{Variation of equation of state parameters with radial coordinate for LMC X-4 with $M = 1.04 M_{\odot}, ~R = 8.3 km$ and $\gamma = 1/3$ in EGB.}
	\label{fig:eospng}
\end{figure}
\begin{figure}[h]
	\centering
	\includegraphics[width=0.5\linewidth]{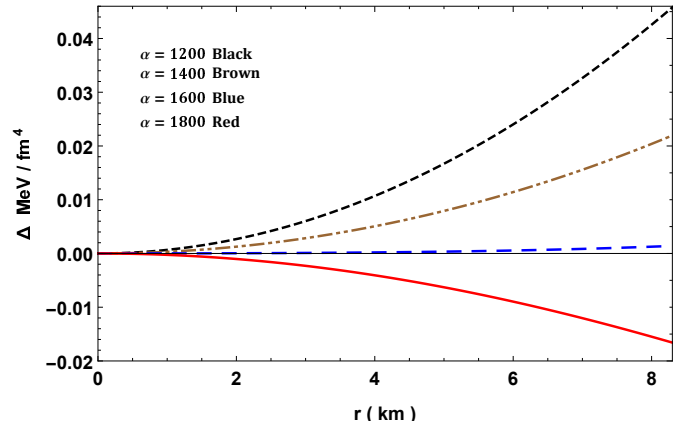}
	\caption[Anisotropy]{Variation of anisotropy with radial coordinate for LMC X-4 with $M = 1.04 M_{\odot}, ~R = 8.3 km$ and $\gamma = 1/3$ in EGB.}
	\label{fig:ani}
\end{figure}
\begin{figure}[h]
	\centering
	\includegraphics[width=0.5\linewidth]{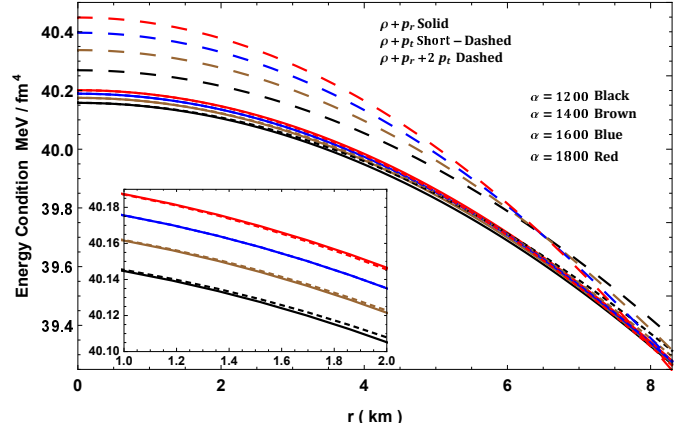}
	\caption[Energy Conditions]{Variation of energy conditions with radial coordinate for LMC X-4 with $M = 1.04 M_{\odot}, ~R = 8.3 km$ and $\gamma = 1/3$ in EGB.}
	\label{fig:ecs}
\end{figure}
\begin{figure}[h]
	\centering
	\includegraphics[width=0.5\linewidth]{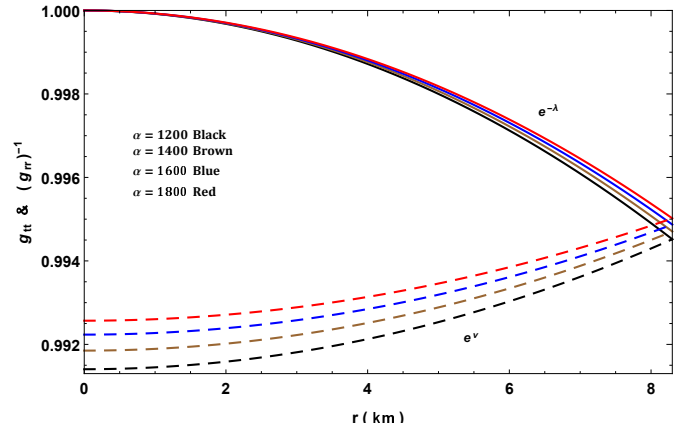}
	\caption[Metric]{Variation of metric potentials with radial coordinate for LMC X-4 with $M = 1.04 M_{\odot}, ~R = 8.3 km$ and $\gamma = 1/3$ in EGB.}
	\label{fig:metr}
\end{figure}

\begin{figure}[h]
	\centering
	\includegraphics[width=0.5\linewidth]{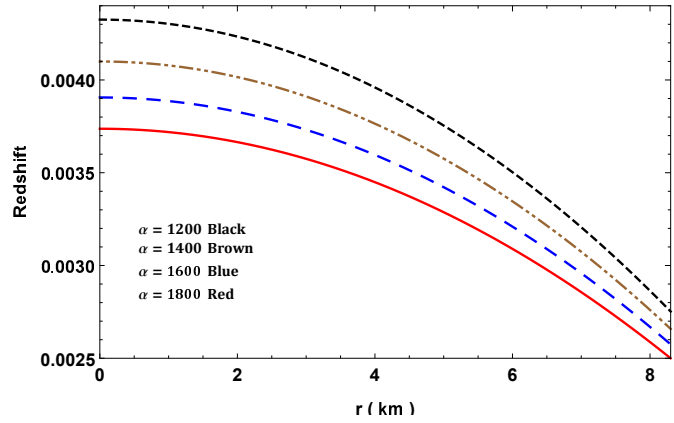}
	\caption[Redshift]{Variation of red-shift with radial coordinate for LMC X-4 with $M = 1.04 M_{\odot}, ~R = 8.3 km$ and $\gamma = 1/3$ in EGB.}
	\label{fig:red}
\end{figure}
\begin{figure}[h]
	\centering
	\includegraphics[width=0.5\linewidth]{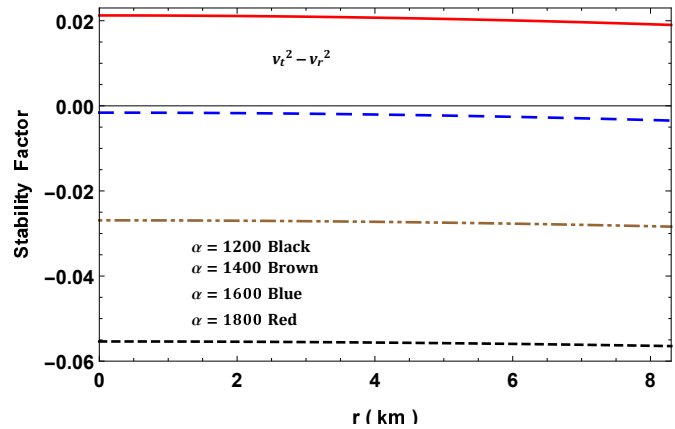}
	\caption[Stability Factor]{Variation of stability factor with radial coordinate for LMC X-4 with $M = 1.04 M_{\odot}, ~R = 8.3 km$ and $\gamma = 1/3$ in EGB.}
	\label{fig:stab}
\end{figure}
\begin{figure}[h]
	\centering
	\includegraphics[width=0.5\linewidth]{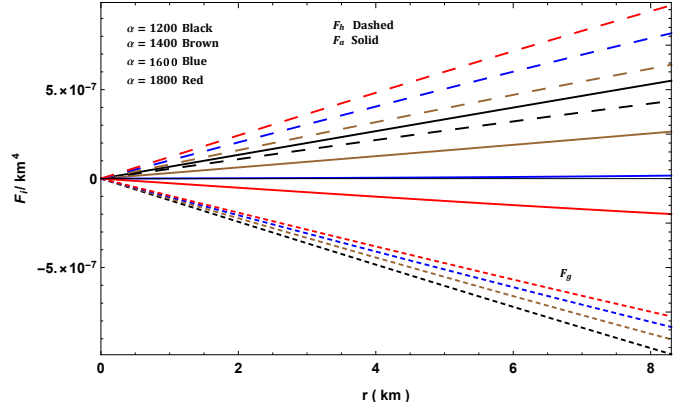}
	\caption[TOV Equation]{Variation of forces in TOV-equation with radial coordinate for LMC X-4 with $M = 1.04 M_{\odot}, ~R = 8.3 km$ and $\gamma = 1/3$ in EGB.}
	\label{fig:tov}
\end{figure}

\begin{figure}[h]
	\centering
	\includegraphics[width=0.5\linewidth]{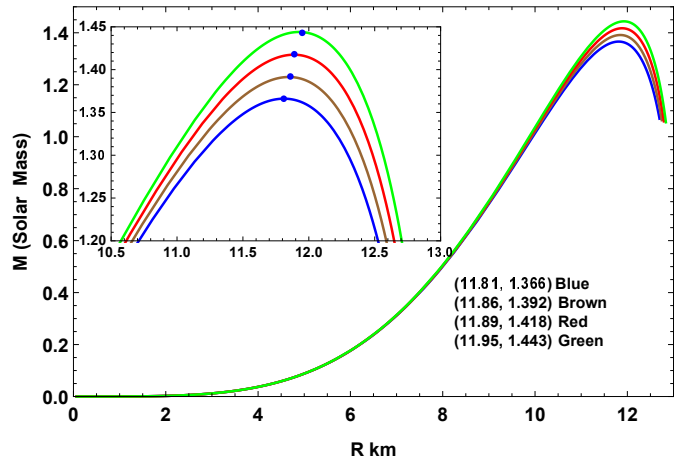}
	\caption[$M-R$ Graph]{$M-R$ graph assuming $M = 1.04 M_{\odot}, ~R = 8.3 km$ in EGB.}
	\label{fig:mr}
\end{figure}

\begin{figure}[h]
	\centering
	\includegraphics[width=0.5\linewidth]{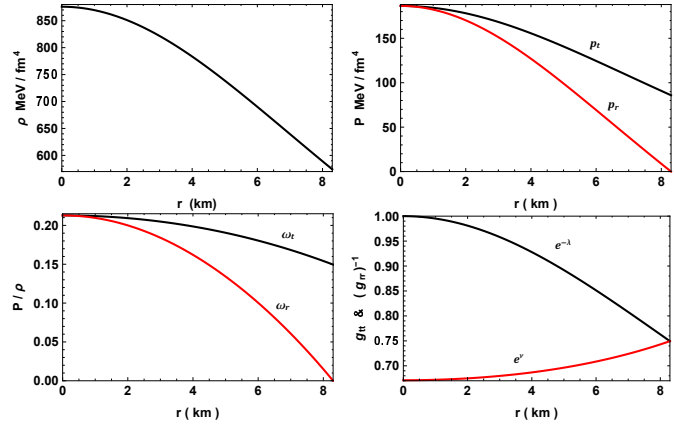}
	\caption{Graphs of metric potentials, density, pressure and $P/\rho$ for LMC X-4 with $M = 1.04 M_{\odot}, ~R = 8.3 km$ and $\gamma = 1/3$ in GR limit $\alpha=0$.}
	\label{fig:gr1}
\end{figure}

\begin{figure}[h]
	\centering
	\includegraphics[width=0.5\linewidth]{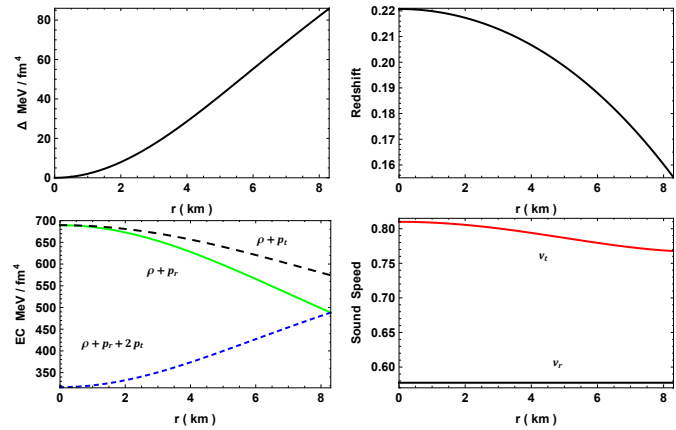}
	\caption{Graphs of anisotropy, red-shift, energy conditions and sound speeds for LMC X-4 with $M = 1.04 M_{\odot}, ~R = 8.3 km$ and $\gamma = 1/3$ in GR limit $\alpha = 0$.}
	\label{fig:gr2}
\end{figure}

\begin{figure}[h]
	\centering
	\includegraphics[width=0.5\linewidth]{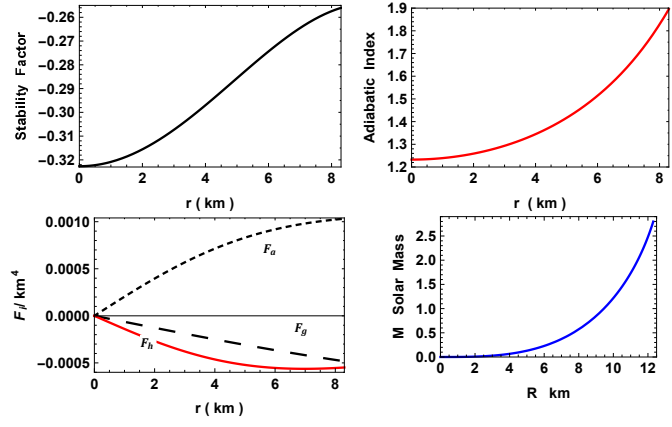}
	\caption{Graphs of stability factor, adiabatic index, TOV-equation and $M-R$ curve with $M = 1.04 M_{\odot}, ~R = 8.3 km$ and $\gamma = 1/3$ in GR limit $\alpha = 0$.}
	\label{fig:gr3}
\end{figure}


\begin{table}[ht]
\centering
\begin{tabular}{|l|l|l|l|l|l|l|l|l|l|}
\hline
Object & ${M \over M_\odot}$ & $R$ & $\gamma$ & $c \times 10^{-3}$ & $\alpha$ & $\xi$ & $\rho_c \times 10^{13}$ & $\rho_b \times 10^{13}$ & $p_c \times 10^{33}$ \\
 & & km & &  & & & $g/cc$ & $g/cc$ & $dyne/cm^2$\\
\hline
LMC X-4	& 1.04 & 8.3 & 0.33 & 0.161 & 1200 & 0.06885 & 7.15 & 6.99 & 8.68\\
SMC X-4 & 1.29 & 8.831 & 0.33 & 0.155 & 1250 & 0.07931 & 6.93 & 6.77 & 9.50\\
EXO 1785-248 & 1.3 & 8.849 & 0.33 & 0.154 & 1300 & 0.12468 & 6.89 &	6.77 & 1.06\\
4U 1820-30 & 1.58 & 9.1 & 0.33 & 0.163 & 1350 & 0.27091 & 7.52 & 7.32 &	1.78\\
PSR J1614-2230 & 1.97 & 9.69 & 0.33 & 0.158 & 1400 & 0.2788 & 7.32 & 7.13 & 1.93\\
\hline
\end{tabular}
\caption{\label{tab1}Parameter of few well-known compact star candidates}
\end{table}

\section*{Results and conclusion}

In this work we have generated the equations governing the dynamical evolution of astrophysical models in the Einstein--Gauss--Bonnet gravity paradigm with anisotropic stresses. After electing to use a strange star equation of state we employed the gravitational potential of Vaidya and Tikekar which was used to construct models of superdense stars in four dimensional gravity. The remaining gravitational potential was settled by solving a differential equation emanating from the equation of state. It was then possible to calculate all the remaining dynamical variables and stability indicators. Graphical plots assisted us to investigate the behavior of the model with and without higher curvature effects. It was found that lower energy densities were realizable for increasing values of the coupling constant $\alpha$. The pulsar LMC X--4 supplied mass and radius values to analyse other features of the star. It was also found that higher curvature terms resulted in a significant reduction in surface gravitational redshift values when compared to the 5 dimensional Einstein star. With regards to stability we concluded that the Gauss--Bonnet terms did not disturb the stability of the model in the Chandrasekhar adiabatic stability sense nor in the sense of the TOV equation  components. Lower sound speeds were evident in the EGB models however neither model became acausal within the radial value. It was shown that the EGB model produced characteristics not out of sync with a range of known compact objects. This study demonstrates that the higher curvature Gauss--Bonnet terms impose a strong influence on the structure of stars and could potentially alter inferences and interpretations of observations of stars at large length scales.

\section*{Acknowledgments}

\noindent SH thanks the National Research Foundation (NRF) of South Africa for financial support through Competitive Grant number CSRP170419227721. LM  is grateful to the NRF for the award of a bursary under the grantholder.

\section*{Author contributions statement}

SH: conceived project, write-up of the paper, exact solution found, analysis of plots. MG: contributed to write-up, conducted detailed analysis of physical properties of model. LM: performed calculations  of dynamical variables and formatting, compiled bibliography. KNS: constructed comprehensive plots of physical quantities, generated table of known compact star data, contributed to the analysis,  and layout of manuscript.



\begin{thebibliography}{99}

\bibitem{amendola} Amendola L. , C. Charmousis and  S. C. Davis, 2007,  {\em JCAP 0710}, {  10} 004
	
\bibitem{abreu}  Abreu H.,  H. Hern\'{a}ndez and L.A. N\'{u}\~{n}ez, 2007, { \em Class. Quant. Grav.}{, 24}, 4631



\bibitem{Batista1}  Batista C. E. M. \textit{et al.}, 2012. {\em  Phys. Rev. D}, {   85}, 084008.
\bibitem{Batista2} Batista C. E. M., J.~C.~Fabris, O.~F.~Piattella and A.~M.~Velasquez-Toribio, 2013. {\em Eur.\ Phys.\ J.\ C} { ,  73}  2425.
\bibitem{beroiz} Beroiz M., G. Dotti and R. J. Gleiser, 2007. {\em Phys. Rev. D} {,  76}, 024012.
	\bibitem{18}  Bogdanos C.,  C. Charmousis ,  B. Goutraux and  R. Zegers, 2009. {\em JHEP} {, 0910}, 37.  		
	\bibitem{3} Boulware D. G. and S. Deser, 1985. { \em Phys. Rev. Lett.} {, 55}, 2656.
	\bibitem{bowers}  Bowers R. L. and  E. P. T. Liang, 1974. {\em Astrophys. J.} {, 188} 657. 	
	\bibitem{buch1}  Buchdahl H. A. 1959. {\em Phys. Rev.}, { 116}, 1027.
	\bibitem{buch2}  Buchdahl H. A. 1984. {\em Class. Quant. Grav.}, { 1}, 301.

	\bibitem{capo}  Caporaso G. and K. Brecher  , 1979. {\em Phys. Rev. D}{, 20}, 1823.   	
	
	\bibitem{mg2} Chan R., L. Herrera and N. O. Santos  , 1993. {\em Mon. Not. R. Astron. Soc.}{, 265}, 533.
	\bibitem{chandra1}   Chandrasekhar S. 1964a. {\em Astrophys. J.}{, 140}, 417.
    \bibitem{chandra2}   Chandrasekhar S. 1964b. {\em Phys. Rev. Lett.}{, 12}, 114.
	\bibitem{chil-hans} Chilambwe B., S.  Hansraj and S. D. Maharaj, 2015. {\em Int. J. Mod. Phys. D}{, 24} 1550051.

	\bibitem{11}  Dadhich N. K.,  A. Molina, A. Khugaev, 2010. {\em Phys. Rev. D}{, 81}, 104026.

 \bibitem{darmois}  Darmois G. J., 1927. {\em Memorial des Sciences Mathematiques}, 25, 1.

	\bibitem{15}  Davis S. C., 2003. {\em Phys. Rev. D}{, 67}, 024030.


\bibitem{dehghani}  Dehghani M. H., 2004. {\em Phys. Rev. D}, {  70}, 064009.
\bibitem{delacruz}  de la Cruz-Dombriz A., P.K.S. Dunsby, and D. Saez-Gomez, 2014. {\em JCAP}, {  2014}, 048 	.
\bibitem{deRham} de Rham C.,  2014. {\em Living Rev. Rel.} {  17}, 7.
\bibitem{doneva} Doneva D. D. and  Yazadjiev S. S., 2021. {\em JCAP} {\bf  2021}, 024.
	\bibitem{duorah}  Duorah H. L. and R. Ray , 1987. {\em Class. Quant. Grav.}{, 4}, 1691.
\bibitem{fabris} Fabris J. C., M. H. Daouda  and O. F. Piattella , 2012.  {\em Phys.\ Lett.\ B} {\bf 711},  232.
	\bibitem{finch}   Finch M. R. and  J. E. F. Skea , 1989. {\em Class. Quant. Grav.}{, 6}, 467.
	
	\bibitem{2} Ghosh S. G., S. Jhingan and D. W. Deshkar , 2014. {\em J. Phys.: Conf. Series}{, 484}, 012013.

\bibitem{goswami} Goswami R., A. Nzioki, S. D. Maharaj and S. G. Ghosh, 2014. {\em Phys. Rev. D} { 90}, 084011.
	\bibitem{mgg}  Govender M.,  A. Maharaj ,  D. Lortan and  D.Day, 2018. {\em Astrophys. Space. Sci.}{, 363}, 165.
	\bibitem{mg4}  Govender M.,  N. Mewalal and  S.Hansraj , 2019. {\em Eur. Phys. J. C}{, 79}, 24.
	\bibitem{m1}  Govender M. and  S. Thirukkanesh , 2015. {\em Astrophys. Space Sci.}{, 358}, 8.  	
	\bibitem{1}  Gross D., 1999. {\em Nucl. Phys. Proc. Suppl.}{, 74}, 426.
\bibitem{vt2} Gupta Y. K. and M. K. Jasim, 2000. {\em Astrophys. Space Sci.} {  272}, 403.

\bibitem{hans-ban1} Hansraj S. and  A. Banerjee, 2018.  {\em Phys. Rev. D} {   97}  104020.
\bibitem{hans-ban2} Hansraj S. and A. Banerjee, 2020. {\em Mod. Phys. Lett. A} { 35} 2050105.
\bibitem{hans-maha}  Hansraj S.,  B. Chilambwe , S. D. Maharaj, 2015. {\em Eur. Phys. J. C }{, 27} 277.
\bibitem{hans-cqg} S. Hansraj, M. Govender, A. Banerjee and N. Mkhize, (2021). {\em Classical and Quantum Gravity}, {  38}, 065018.
\bibitem{Heydarzade}   Y. Heydarzade, H. Moradpour and F. Darabi, 2017. {\em  Can.J.Phys.},  {  95}, 1253.
\bibitem{herrera}  Herrera L., 1992. {\em Phys. Lett. A}{, 165}, 206.
	\bibitem{mg1} Herrera L., G. Le Denmat and N.O. Santos, 1989. {\em Mon. Not. R. Astron. Soc.}{, 237}, 257.
    \bibitem{mg3} Herrera L., G. Le Denmat and N.O. Santos   , 2012. {\em Gen. Rel. Gravit.}{, 44}, 1143.
   	\bibitem{horn} Horndeski G. W., 1974. {\em Int. J. Theor. Phys.}{, 10},  363.


\bibitem{israel} Israel W.,(1966). {\em Nuovo Cim. B}, { 44}, 1.

   	\bibitem{10}  Jhingan S. and  S. G. Ghosh, 2010. {\em Phys. Rev. D}{, 81}, 024010.
\bibitem{kaluza}  Kaluza T., 1921. {\em  Sitz. Ber. Preuss. Akad. Wiss.},  966
	\bibitem{14}  Kang Z.,  Y. Zhan--Ying , Z. De--Cheng and  Y. Rui--Hong , 2012. {\em Chin. Phys. B}{,  21}  020401.
	\bibitem{kamarkar}   Karmarkar K.R., 1948. {\em Proc. Ind. Acad. Sci. A}{, 27}, 56.
\bibitem{kastor}  Kastor D.,(2013). {\em Class. Quantum Grav.} { , 30} 195006.

\bibitem{klein}  Klein O., (1926). {\em  Zeit. f. Physik} {, 37},  895.

    \bibitem{kohler}  Kohler M. and K.L.  Chao, 1965. {\em Z. Naturforsch. Ser. A}{,  20}, 1537. 	  		
	\bibitem{mous}  Koliogiannis P. S. and  C. C. Moustakidis, 2019. {\em  Astrophys Space Sci}{, 364},  52.
	\bibitem{vt1} Kumar J., A. K.Prasad , S. K. Maurya and A. Banerjee , 2018. {\em Eur. Phys. J. C} { , 78}, 540.

\bibitem{love1}  Lovelock D., (1971). {\em J. Math. Phys.} {, 12}, 498.
\bibitem{love2}  Lovelock D., (1972). {\em J. Math. Phys.} {,13}, 874.


\bibitem{maartens} Maartens R. and  K. Koyama, 2010. {\em Living Reviews in Relativity}, { 13}, 10.
	\bibitem{22} Mafa Takisa P. and S. D. Maharaj , 2013. {\em Astrophys. Space. Sci.}{, 343}, 569.
	\bibitem{9} Maeda H., 2006. {\em Phys. Rev. D}{, 73}, 104004.
\bibitem{maha-hans}  Maharaj S. D., B. Chilambwe , S. Hansraj , 2015.  {\em Phys. Rev. D}{, 91},  084049. 	
    \bibitem{23} Maharaj  S. D.,  J. Sunzu and  S. Ray , 2014. {\em Eur. Phys. J. Plus}{, 129}, 3.
    \bibitem{maurya-hans}  Maurya S.K.,   A. Banerjee and  S. Hansraj, 2018. {\em Phys. Rev. D}{, 97}, 044022.
   \bibitem{m4}  Maurya S. K., A. Banerjee,  M. K. Jasim ,  J. Kumar , A. K. Prasad and  A. Pradhan , 2019. {\em Phys. Rev D}{, 99} 044029.
    \bibitem{m3}  Maurya S. K. and   M. Govender, 2017. {\em Eur. Phys. J. C}{, 77}, 420.
	\bibitem{m2}  Maurya S. K., B. S. Ratanpal and M. Govender, 2017. {\em Annals of Phys.}{, 382}, 36.
	\bibitem{molina} Molina A.,  N. K. Dadhich and A. Khugaev  , 2017. {\em Gen. Rel. Grav.}{, 49}, 96.
	\bibitem{5} Myers R. C. and  M. J. Perry , 1986. {\em Ann. Phys.}{, 172}, 304.  	
	\bibitem{8}  Myers R. C. and J. Z. Simons , 1988. {\em Phys. Rev. D}{, 38}, 2434. 		
\bibitem{navarro}  Navarro. A. and J. Navarro, (2011). {\em  Journal of Mathematical Physics} {, 61}, 1950.


\bibitem{nils1}  Nilsson U.S. and C. Uggla, (2001). {\em Ann. Phys.} , 286, 278.
\bibitem{nils2}  Nilsson U.S. and C. Uggla, (2001). {\em Ann. Phys.} , 286, 292.


\bibitem{pano} Panotopoulos G and Rinc$\acute{o}$n $\acute{A}$., 2019.  {\em Eur.Phys.J. Plus}, { 134}, 9.
\bibitem{vt3} Paul, B. C., P. K. Chattopadhyay , S. Kamarkar and R. Tikekar , 2011. {\em Mod. Phys. Lett. A} {, 26}, 575.
	\bibitem{perl}  Perlmutter S. {\it{et al.}}, 1999. {\em  Astrophys. J.}{, 517} 565.
\bibitem{resco} Resco  M. A., A de la Cruz-Dombriz, F. J. Lianes-Esstrada and V. Z. Castrillo, (2016).  {\em Phys. Dark Univ.}, {   13}, 147.
 \bibitem{Rastall}  Rastall P., (1972).  {\em Phys. Rev. D}, { 6}, 3357.
\bibitem{Rastall1}  Rastall  P., (1976). { \em Can. J. Phys.}, {  54}, 66.

	\bibitem{riess} Riess A. G. {\it{et al.}}, 1998.  {\em  Astron. J.}{, 116}, 1009.	
\bibitem{saslaw}  Saslaw W.C., S.D. Maharaj and N.K. Dadhich,(1996). {\em Astrophys. J.} {, 471}, 571.


\bibitem{senovilla}  Senovilla J. M. M.,(2013). {\em Phys. Rev. D} {, 88}, 064015.
\bibitem{vt4} Sharma R., S. Das , M. Govender and D. M. Pandya, 2020. {\em Ann. Phys.} {, 414} 168079.

\bibitem{silva} Silva G. F., O. F. Piattella, J. C. Fabris, L. Casarini  and  T. O. Barbosa, 2013. {\em Grav. Cosmol.} {, 19},  156.

\bibitem{soti} Sotiriou T. P. and V. Faraoni , 2010, {\em Reviews of Modern Physics} {, 82} 451.
\bibitem{staro} Starobinsky A.A., 1980. {\em Phys. Lett. B} {,  91}, 99.

		
	\bibitem{4} Tangherlini F. R., 1963. {\em Il Nuovo Cimento}{, 27}, 636.
	\bibitem{7} Torii  T. and  H. Maeda, 2005. {\em Phys. Rev. D}{, 71}, 124002.
	
	\bibitem{vt}  Vaidya P.C. and R.  Tikekar , 1982. { \em Journal of Astrophysics and Astronomy}{, 3}, 325.
	\bibitem{velay} Velay-Vitow J. and A. DeBenedictis, (2017). {\em Phys. Rev. D} {,  96}, 024055.

	\bibitem{wal}  Walecka J. D., 1975. {\em Phys. Lett.}{, 59}, 109.
	\bibitem{6} Wheeler J. T., 1986. {\em  Nucl. Phys. B}{, 268},  737. 	
	\bibitem{wright}  Wright M., 2016. {\em Gen. Rel. Gravit.}{, 48},  93.

	
	\bibitem{xian} Xian-Feng Z. and J. Huan-Yu, 2014. {\em Revista Mexicana de Astronomia y Astrofisica}{, 50}, 103.
	
	


	
\end{thebibliography}
\end{document}